\documentclass{article}
\usepackage[utf8]{inputenc}

\usepackage{amssymb}
\usepackage{color,colortbl}
\usepackage{epstopdf}
\usepackage{graphicx}
\usepackage{tabularx}
\usepackage{booktabs}
\usepackage{amsmath}
\DeclareMathOperator{\tr}{\rm tr\,}
\usepackage{dsfont}
\usepackage{xstring}
\RequirePackage[colorlinks,citecolor=blue,urlcolor=blue,linkcolor=blue]{hyperref}

\usepackage[normalem]{ulem}
\usepackage{arydshln}

\usepackage[capitalize]{cleveref}
\Crefname{section}{Section}{Sections}
\Crefname{figure}{Figure}{Figures}
\Crefname{table}{Table}{Tables}
\Crefname{appendix}{Appendix}{Appendices}
\Crefname{equation}{Eq.}{Eqs.}
\newcommand{\Onlinecite}[1]{%
    \IfSubStr{#1}{,}{Refs}{Ref}.~\cite{#1}%

}

\title {Unveiling SU(3) Flux Tubes At Nonzero Temperature: \\Electric Fields and Magnetic Currents  \\ }

\author{
    Marshall Baker
    \footnote{email: mbaker4@uw.edu}
    \\ 
    Department of Physics, University of Washington, 
    \\ WA 98105 Seattle, USA
\and
    Volodymyr Chelnokov
    \footnote{email: chelnokov@itp.uni-frankfurt.de}
    \\
    Institut f\"ur Theoretische Physik, Goethe Universit\"at, 
    \\ 60438 Frankfurt am Main, Germany
\and
    Leonardo Cosmai
    \footnote{email: leonardo.cosmai@ba.infn.it}
    \\
    INFN - Sezione di Bari, I-70126 Bari, Italy
\and
    Francesca Cuteri
    \footnote{email: cuteri@itp.uni-frankfurt.de}
    \\
    Institut f\"ur Theoretische Physik, Goethe Universit\"at, 
    \\ 60438 Frankfurt am Main, Germany
\and
    Alessandro Papa
    \footnote{email: alessandro.papa@fis.unical.it}
    \\
    Dipartimento di Fisica, Universit\`a della Calabria
    \\ and INFN - Gruppo collegato di Cosenza, 
    \\ I-87036 Arcavacata di Rende, Cosenza, Italy
}

\definecolor{Gray}{gray}{0.9}

\begin{document}

\maketitle

\begin{abstract}
We report on the results of measuring the chromoelectric fields in 
a flux tube created by a static quark-antiquark pair in the finite-temperature SU(3) gauge theory. Below the deconfinement temperature the field behavior is similar to the zero-temperature case. Above the deconfinement temperature the field shape remains the same, but the field
values drop when the distance between quark and antiquark increases, thus showing the disappearance of
confining potential. 
\end{abstract}

\section{Introduction}
In previous papers \cite{Baker:2018mhw,Baker:2019gsi,Baker:2022cwb} we have carried out lattice simulations of the color fields in the region between a static quark and anti-quark in pure $\text{SU(3)}$ gauge theory, and extracted the gauge-invariant non-perturbative longitudinal electric fields $\vec{E}^\text{NP}$ and magnetic currents $\vec{J}_\text{mag}$. 

In this paper we carry out a corresponding analysis of $\text{SU(3)}$ flux tubes at nonzero temperature, both above and below the deconfinement temperature $T_\text{c}$.
In particular we examine the behavior of $\vec{E}^\text{NP}$ and $\vec{J}_\text{mag}$ in the midplane between the quark and the anti-quark.

We discuss the use of the lattice operator $\rho^\text{conn}_{W, \mu \nu}$, the connected correlator involving Wilson loops with the largest possible extension in the temporal direction,  in order
to measure flux-tube fields at nonzero temperatures. Below $T_c$ we compare the nonperturbative fields extracted with the use of the Wilson loop correlator $\rho^\text{conn}_{W, \mu \nu}$ with those extracted using the Polyakov loop correlators, $\rho^\text{conn}_{P, \mu \nu}$.

The force on the magnetic currents in nonzero temperature flux tubes has the same form $\vec{f} = \vec{J}_\text{mag} \times \vec{E}^\text{NP}$ as in the $T = 0$ flux tubes, where $\vec{J}_\text{mag}$ and $\vec{E}^\text{NP}$ are measured 
 by lattice simulations at nonzero temperature. This is discussed in Section~3.

Section 4 describes our lattice setup and the smearing procedure used to extract physical information from our simulations. Our numerical results are presented in Section 5 and the conclusions given in Section 6.

\section{Connected correlator, field strength tensor, comparison of Polyakov loop correlator and Wilson loop}

\begin {figure}[htb]
\centering
\includegraphics[width=0.7\linewidth,clip]{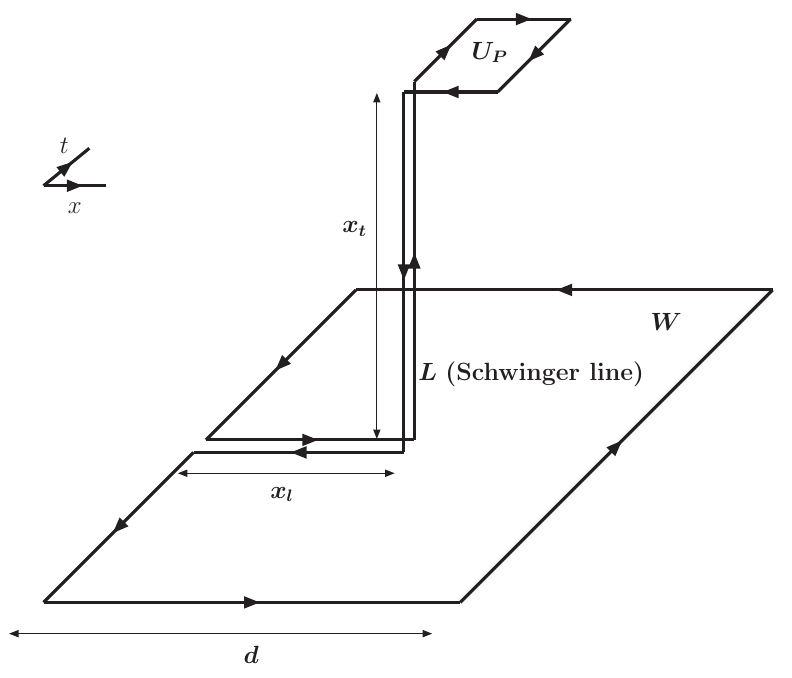}
\includegraphics[width=0.25\textwidth,clip]{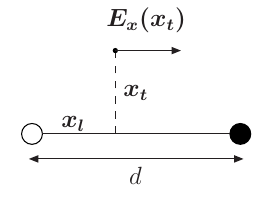}
\caption{The connected correlator between the plaquette $U_{P}$ and the Wilson loop (subtraction in $\rho_{W,\,\mu\nu}^\text{conn}$ not explicitly drawn).
The longitudinal electric field $E_x(x_t)$ at some fixed displacement $x_l$ along the axis connecting the static sources (represented by the white and black circles), for a given value of the transverse distance $x_t$.}               \label{fig:op_W}
\end{figure}

At zero temperature, the spatial distributions of the color fields induced by a static quark-antiquark pair can be obtained from lattice measurements of the connected correlation function $\rho^\text{conn}_{W, \mu \nu}$~\cite{DiGiacomo:1989yp} of a plaquette $U_P = U_{\mu \nu} (x)$ in the ${\mu \nu}$ plane, and a Wilson loop $W$ (see Fig.~\ref{fig:op_W}),
\begin{equation}
    \rho^\text{conn}_{W, \mu \nu} = \frac {\langle\tr (WLU_PL^*)\rangle}{\langle\tr(W)\rangle} - \frac{1}{N} \frac {\langle\tr (U_P) \tr (W)\rangle}{\langle\tr(W)\rangle}\;,
    \label{connected1}
\end{equation}
$N=3$ being the number of QCD colors. The correlator $\rho^\text{conn}_{W, \mu \nu}$ provides a lattice definition of a gauge-invariant field strength tensor $\langle F_{\mu \nu}\rangle_{q \bar{q}} \equiv F_{\mu \nu}$ carrying a unit of octet charge, while possessing the space-time symmetry properties of the Maxwell field tensor of electrodynamics, 
\begin{equation}
\rho^\text{conn}_{W, \mu \nu} \equiv~~ a^2 g\langle F_{\mu \nu}\rangle_{q \bar{q}} ~~ \equiv~~ a^2 g ~F_{\mu \nu}\;.
\label{connected2}
\end{equation}

When the plaquette $U_P$ lies in the $\hat 4 \hat 1$ plane, the measured $\hat 4 \hat 1$ component of the field tensor determines $E_x$, the component of the electric field along the $q\bar{q}$ axis $ E_x = F_{41}$; {\em i.e.}, the longitudinal component of the electric field at the position corresponding to the center of the plaquette. 

When $U_P$ is in the $\hat 4 \hat 2$ plane, $F_{42} = E_y$, a component of the electric field transverse to the $q\bar{q}$ axis,
when $U_P$ is in the $\hat 2 \hat 3$ plane, $F_{23} = B_x$,  the  longitudinal component of the magnetic field, {\it etc}.

We adopted the lattice operator $\rho^\text{conn}_{W, \mu \nu}$ in our previous studies
in $\text{SU(3)}$ pure gauge theory~\cite{Baker:2018mhw,Baker:2019gsi,Baker:2022cwb}, for distances $d$ ranging from $0.37$~fm to $1.25$~fm, and found that all components $B_i $ of the magnetic field were equal to zero within statistical errors.

\begin {figure}[htb]
\centering
\includegraphics[width=0.9\linewidth,clip]{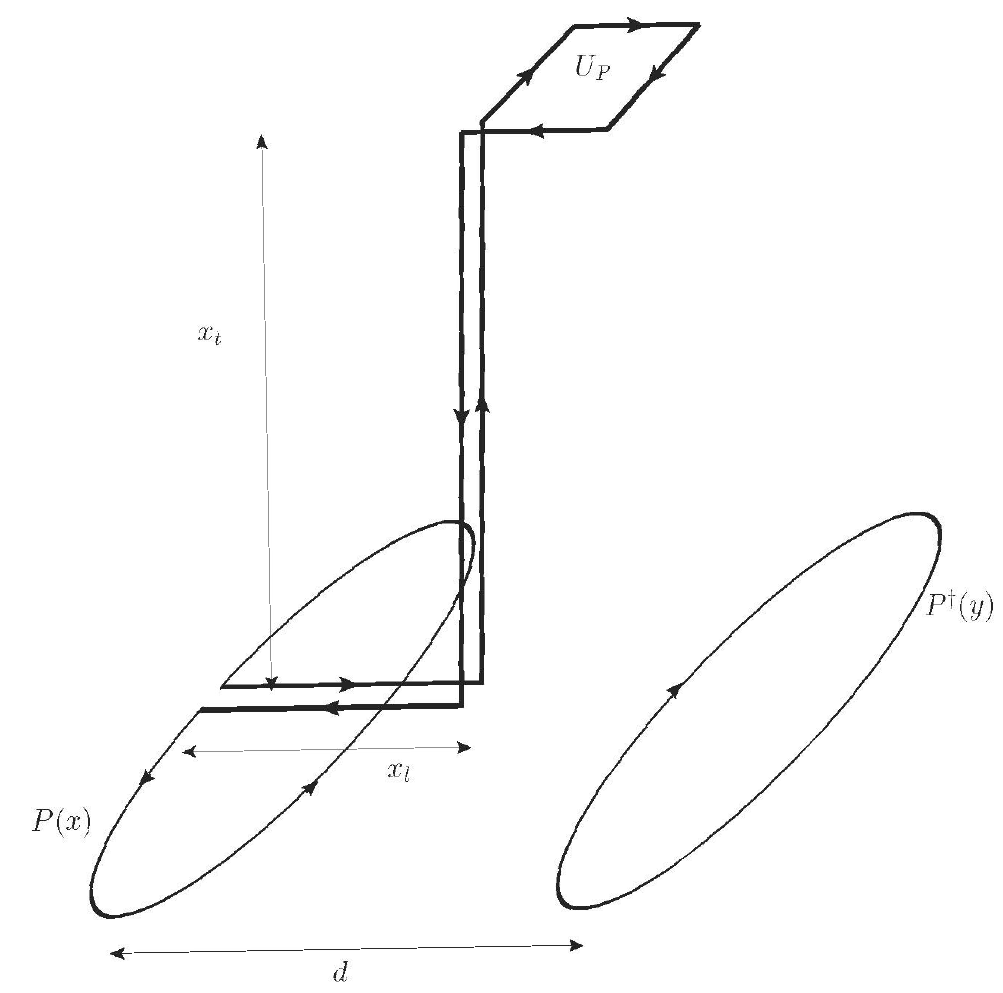}
\caption{The connected correlator between the plaquette $U_{P}$ and the Polyakov loop (subtraction in $\rho_{P,\,\mu\nu}^\text{conn}$ not explicitly drawn).}  \label{fig:op_P}
\end{figure}

At nonzero temperatures, the operator $\rho^\text{conn}_{W, \mu \nu}$ should be replaced
by~\cite{Skala:1996ar} (see Fig.~\ref{fig:op_P})
\begin{equation}
    \rho^\text{conn}_{P, \mu \nu} = \frac {\langle\tr (PLU_PL^*)\tr(P^\dagger)\rangle}{\langle\tr(P)\tr(P^\dagger)\rangle} - \frac{1}{N} \frac {\langle\tr (U_P) \tr (P)\tr(P^\dagger)\rangle}{\langle\tr(P)\tr(P^\dagger)\rangle}\;,
    \label{connected1P}
\end{equation}
where the $P$ and $P^\dagger$ denote two parallel Polyakov lines with opposite orientations, separated by the distance $d$. The lattice operator $\rho^\text{conn}_{P, \mu \nu}$ was adopted in some previous studies of $\text{SU(3)}$ pure gauge theory at zero temperature~\cite{Cea:2013oba,Cea:2014uja,Cea:2014hma} and at nonzero temperature across the phase transition~\cite{Cea:2015wjd}. In those works, lattice measurements were limited to the longitudinal chromoelectric field and were carried out on the transverse plane at the midpoint between the sources, {\it i.e.} for $x_l=d/2$.

Lattice measurements based on the use of the lattice operator $\rho^\text{conn}_{P, \mu \nu}$
for longitudinal distances $x_l$ ranging from zero to $d$ show that,
in the $\text{SU(3)}$ deconfined phase, the chromoelectric field is much larger for $x_l$ near zero than for $x_l$ near $d$. This asymmetric behavior can be explained by the effect of renormalization, which is dependent on the length and shape of the Schwinger line (see~\cite{Battelli:2019lkz}). Performing
lattice measurements on {\em smeared} Monte Carlo ensembles is an effective way to take into account these renormalization effects. 

It turns out, however, that even when $\rho^\text{conn}_{P, \mu \nu}$ is measured on smeared ensembles, a strong asymmetry in $x_l$ survives, which can be explained by the inefficacy of the smearing procedure near the sources. This problem does not occur in the low-temperature phase of $\text{SU(3)}$.

For this reason, we decided to adopt, instead of $\rho^\text{conn}_{P, \mu \nu}$, the connected correlator $\rho^\text{conn}_{W_{\rm max}, \mu \nu}$, involving {\em maximal} Wilson loops, {\it i.e.} loops with the largest possible extension in the temporal direction.
The operators $\rho^\text{conn}_{W_{\rm max}, \mu \nu}$ and $\rho^\text{conn}_{P, \mu \nu}$
differ in a sense which can be understood considering their counterparts when the Schwinger line and the plaquette are removed.

As shown in Ref.~\cite{Jahn:2004qr}, a standalone maximal Wilson loop with a given extension $d$ in the spatial direction is equivalent to the singlet correlator of two Polyakov loops, $\langle {\rm tr}(P P^\dagger)\rangle$, at distance $d$, calculated in the axial gauge, on a lattice with periodic boundary conditions, and gives
access to the static potential for a quark-antiquark pair in the singlet state. On
the other hand, a Polyakov loop correlator of the form $\langle {\rm tr}(P){\rm tr}(P^\dagger)\rangle$ gives access to a combination of the quark-antiquark static potential in the singlet and in the octet states.

\begin{figure}[htb]
\centering
\includegraphics[width=0.9\linewidth]{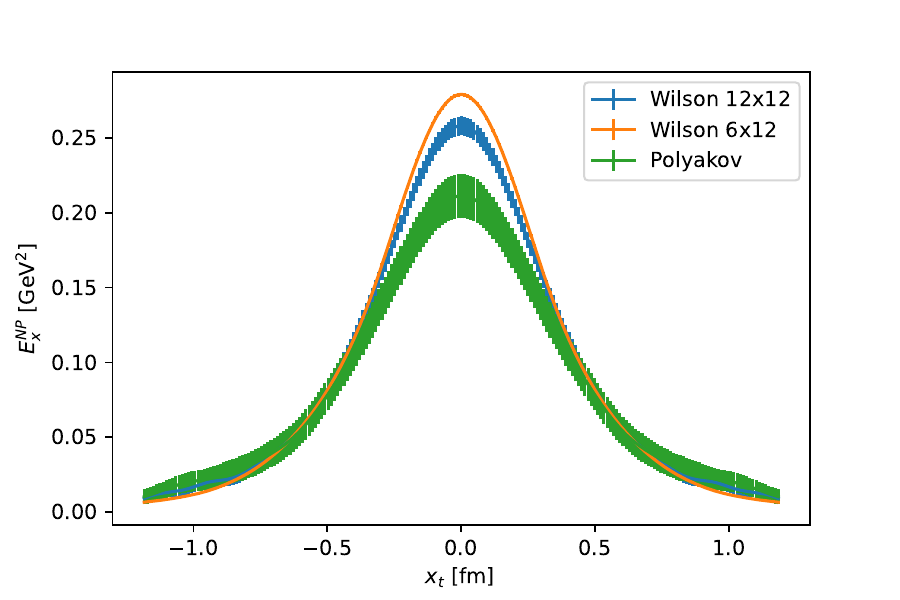}
\caption{Comparison of the nonperturbative field extracted using the Polyakov loop correlator, and Wilson loop correlators with temporal size $N_T=12$ and $N_T/2=6$, at the transverse midplane $x_l = d / 2$ for $d=0.842$~fm and $T=0.8 T_\text{c}$.}  \label{fig:correlator_comparison}
\end{figure}

 To quantify the difference between $\rho^{\rm conn}_{W_{\rm max}, \mu \nu}$ and $\rho^\text{conn}_{P, \mu \nu}$, in Fig.~\ref{fig:correlator_comparison} we present a comparison of the nonperturbative part of the chromoelectric field  on the midplane  between the static sources, as extracted from $\rho^\text{conn}_{P, \mu \nu}$, from
$\rho^\text{conn}_{W_{\rm max}, \mu \nu}$, at
$T=0.8 T_\text{c}$ corresponding to $N_T=12$. 

In the same figure we show also the field determination based on a nonmaximal Wilson loop, with an extension in the time direction equal to six lattice spacings and therefore half-sized with respect to the maximal one; it is interesting to observe that it gives a nonperturbative field exceeding that from the maximal Wilson loop. 

\section{The `Maxwell' picture of the Yang-Mills flux tube: Magnetic Currents} 

The possible presence of magnetic currents in SU(3) lattice gauge theory theory was pointed out in Ref.~\cite{Skala:1996ar}, where it was noted that, in contrast to the magnetic monopoles in U(1) lattice gauge theory, the magnetic currents in non-Abelian lattice gauge theory need not be quantized.

In our previous paper~\cite{Baker:2019gsi} we showed that, using the analogy with the basic principles of electrodynamics, the string tension $\sigma$ can be obtained from the stress tensor $T_{\alpha \beta}$,

\begin{equation}
 T_{\alpha \beta} = F_{\alpha \lambda} F_{\beta \lambda} - \frac{1}{4} \delta_{\alpha \beta} F_{\mu \lambda} F_{\mu \lambda}\ , 
 \label{stress}   
\end{equation}
where $F_{\mu \nu}$  is the Maxwell-like field,  Eq.~\eqref{connected2}, measured in our simulations of the connected correlator.

The Maxwell picture of the Yang-Mills flux tube emerges from use of the divergence of  the  stress tensor $T_{\alpha \beta}$ to calculate the force density $f_\beta$,
\begin{equation}
    f_\beta =  \frac{\partial}{\partial x^\alpha} T_{\alpha \beta}\ ,
    \label{Fs}
\end{equation}
 without requiring that the field tensor $F_{\mu \nu}$ satisfy Maxwell's equations.
 
Eq.~\eqref{Fs} yields $f_\beta$ as the magnetic  Lorentz force density acting on a magnetic current density $J_\alpha^\text{mag}$ that circulates about the axis of the flux tube, 
\begin{align}
     f_\beta  =  - F_{\mu \lambda} ~\frac{1}{2} \epsilon_{\alpha \beta\mu \lambda} J^\text{mag}_\alpha \ , 
    \label{fdensity}
\end{align}
where
 \begin{equation}
     J_\alpha^{\rm mag} \equiv \frac{1}{2} \epsilon_{\alpha \beta\mu \lambda} \frac{\partial F_{\mu \lambda}}{\partial x^\beta},~~~~~~~( \epsilon_{4123} = 1) \;.
     \label{Jmag2}
 \end{equation}

Using the field tensor $F_{\mu \nu}$ measured in our simulations of the connected correlator at nonzero temperatures  to evaluate Eq.~\eqref{Jmag2} yields the magnetic current density in a flux tube at nonzero temperature.

Eq.~\eqref{Jmag2} determines the spatial components of the magnetic current density $J^\text{mag}_i$,~~$i = 1,2,3$ in terms of  the electric components of the field tensor, ${F_{4 k} =E_k}$.  The magnetic components of the field tensor, $\frac{1}{2} \epsilon_{ijk} F_{jk}$,    vanish. 

Then Eq.~(\ref{Jmag2})  becomes
\begin{equation}
  \vec{J}_\text{mag} = \vec{\nabla} \times \vec{E} 
    \label{fieldderivs1}
 \end{equation} 
and Eq.~(\ref{fdensity}) becomes
\begin{equation}
    \vec{f} =  \vec{J}_\text{mag}  \times \vec{E}\ .
    \end{equation}
    
Replacing the color electric field $\vec{E}$  by its non-perturbative longitudinal component $\vec{E}^\text{NP}$ yields
\begin{equation}   
  \vec{f} \xrightarrow{} \vec{J}_{\rm mag}  \times \vec{E}^\text{NP},
\label{vecfdensity}
\end{equation}
the confining force density $\vec{f}$ directed toward the flux-tube axis.

We now calculate the confining force $\vec{F}$. 

We imagine cutting the flux tube along any plane containing its axis.
Integrating the force density $\vec{f}$ (Eq.~(\ref{vecfdensity})) over one half of the cut flux tube yields the force $\vec{F}$ on that half; integrating $\vec{f}$ over the other half of the tube yields an equal and opposite force on that half, pushing the two halves together. The resulting `squeezing' force $\vec{F}$  then confines the flux tube in the transverse direction.

This is the `Maxwell' picture of confinement.

\section{Lattice setup and smearing procedure}
We measured the color fields, as defined in Eq.~(\ref{connected1}), generated by a quark-antiquark pair separated by a distance $d$. 
 We set the physical scale for the lattice spacing according to Ref.~\cite{Necco:2001xg}:
\begin{align}
\label{NSscale}
& a(\beta) = r_0 \! \times \! \exp\left[c_0+c_1(\beta\!-\!6)+c_2(\beta\!-\!6)^2+c_3(\beta\!-\!6)^3\right], \nonumber\\
& r_0 = 0.5 \; {\rm fm}, \nonumber\\
& c_0=-1.6804 \,,  c_1=-1.7331 \,,\nonumber\\
& c_2=0.7849 \,, c_3=-0.4428 \,,
\end{align}
for all $\beta$ values in the range $5.7 \le \beta \le 6.92$.
In this scheme, the value of the square root of the string tension is $\sqrt{\sigma} \approx 0.465 \, {\textrm{GeV}}$
(see Eq.~(3.5) in Ref.~\cite{Necco:2001xg}).

The correspondence between $\beta$ and the distance $d$, shown in
Table~\ref{betavalues}, was obtained from this parameterization. 
We performed measurements in the temperature range $0.8 \le T/T_\text{c} \le 2.0$.
The distance in lattice units between quark and antiquark corresponds to the spatial size of the Wilson loop in the connected
correlator of Eq.~\eqref{connected1}.

  The connected correlator defined in Eq.~(\ref{connected1}) exhibits large
  fluctuations at the scale of the lattice spacing, which are responsible
  for a bad signal-to-noise ratio. To extract the physical information carried
  by fluctuations at the physical scale (and, therefore, at large distances
  in lattice units) we smoothed out configurations by a {\em smearing}
  procedure. Our setup consisted of (just) one step of HYP
  smearing~\cite{Hasenfratz:2001hp} on the temporal links, with smearing
  parameters $(\alpha_1,\alpha_2,\alpha_3) = (1.0, 0.5, 0.5)$, and
  $N_{\rm HYP3d}$ steps of HYP3d smearing~\cite{Hasenfratz:2001hp} on the spatial links,
  with smearing parameters $(\alpha_1,\alpha_3) = (0.75,0.3)$.
  $N_{\rm HYP3d}$ is chosen separately for each observable 
  in a way that maximizes the signal value, as described in \cite{Baker:2022cwb}.
  In Table~\ref{betavalues} we summarize our numerical simulations.
 
\begin{table}[tb]
\setlength{\tabcolsep}{0.75pc}
\centering
\caption[]{Summary of the numerical simulations.}
\begin{tabular}{ccccccc}
\toprule
lattice & $\beta$ & $a(\beta)$ [fm]  & $d/a$ & $d$ [fm] & $T/T_\text{c}$  & statistics\\
\midrule
$48^3 \times 12$ & 6.100 & 0.0789097 & 12 & 0.946917 & 0.8 & 2400 \\
$48^3 \times 12$ & 6.381 & 0.052633 & 12 & 0.631597 & 1.2 & 340 \\
$48^3 \times 12$ & 6.381 & 0.052633 & 16 & 0.842129 & 1.2 & 1500 \\
$48^3 \times 12$ & 6.544  & 0.0420845 & 15 & 0.631267 & 1.5 & 1100 \\
$32^3 \times 8$ & 6.248  & 0.0631757 & 10 & 0.631757 & 1.5 & 2580 \\
$48^3 \times 12$ & 6.778  & 0.0315769 & 20 & 0.631537 & 2.0 & 1020 \\
\bottomrule
\end{tabular}
\label{betavalues}
\end{table}

\section{Numerical results}
\subsection{Scaling check}

\begin {figure}[htb]
\centering
\includegraphics[width=0.68\linewidth]{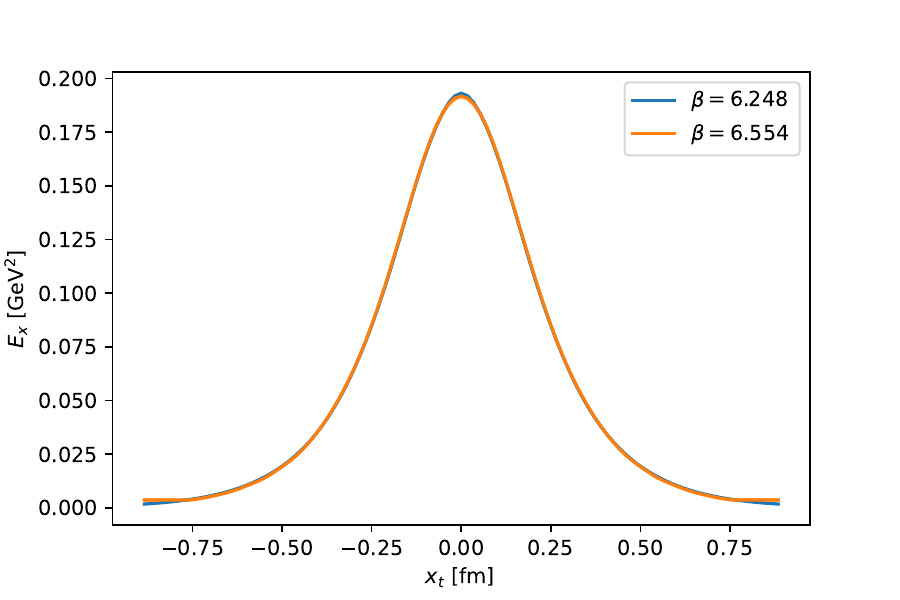} \\
\includegraphics[width=0.68\linewidth]{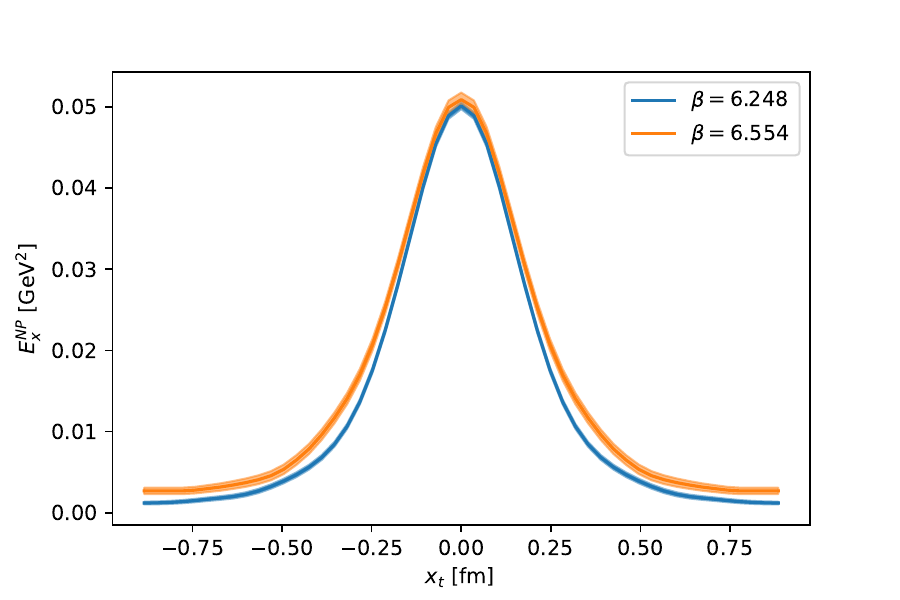} \\
\includegraphics[width=0.68\linewidth]{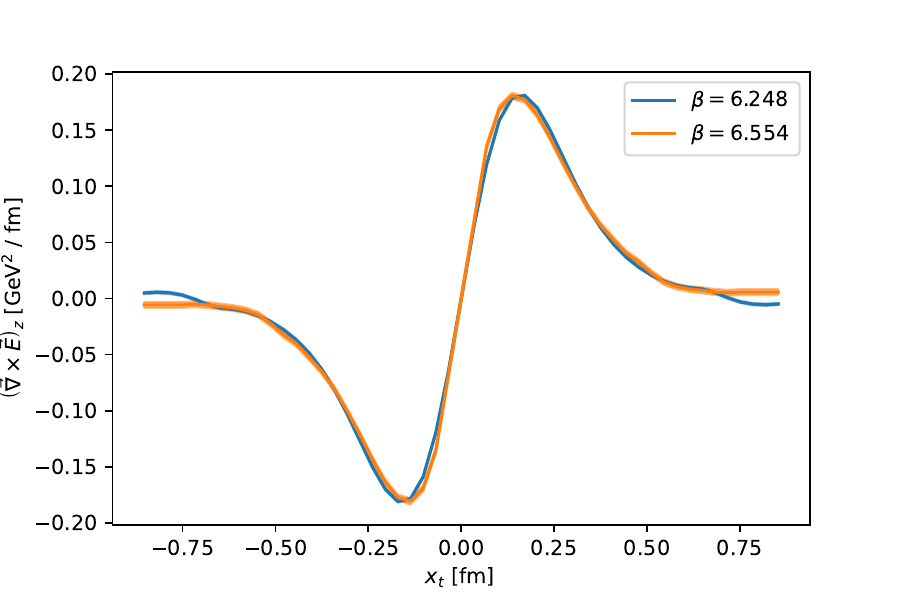}
\caption{Scaling analysis of (from top to bottom), full longitudinal chromoelectric field, nonperturbative chromoelectric field, and chromomagnetic current density.
Comparison is done at $T=1.5 T_\text{c}$ for the fields and current at the midplane with 
$\Lambda = 48^3 \times 12$ , $\beta = 6.544$ , $a \approx 0.042 $~fm, and $\Lambda = 32^3 \times 8$, $\beta = 6.248$, $a \approx 0.063$~fm.}
\label{fig:scaling-check}
\end{figure}

To make sure that we are close enough to the continuum limit, we performed 
a scaling check, comparing the fields and current at the midplane for two parameter sets having different lattice step size $a$ ($0.063$~fm and $0.042$~fm) and different distance between quark and antiquark in lattice units ($10 a$ and $15 a$), but the same temperature $T=1.5 T_\text{c}$ and physical quark-antiquark separation $d \approx 0.631$~fm. The results are shown in \cref{fig:scaling-check}. To be able to compare results exactly at the midplane and avoid the discrepancy due to slightly different location of the points at which the fields are measured, a spline interpolation of the field values at the discrete lattice points was employed. 

The discrepancy between the full field values does not exceed $2 \cdot 10^{-3} \mathrm{\ GeV}^2$, and in most of the cases lies within the error bounds. For the nonperturbative field the discrepancy reaches $3.5 \cdot 10^{-3} \mathrm{\ GeV}^2$ -- up to $4.5 \sigma$, and is much more visible in \cref{fig:scaling-check}, due to the low value of the nonperturbative field itself. The discrepancy in the current density reaches $1.6 \cdot 10^{-2} \mathrm{\ GeV}^2 / \mathrm{fm}$ -- about $5 \sigma$.

This shows that the raw data extracted from the lattice have a negligible contribution 
from finite lattice step (compared to the stochastic errors), though the analysis 
and extraction of derived quantities may introduce discrepancies equal to several standard stochastic errors. 

\subsection{3d plots and asymmetry}

\begin {figure}[htb]
\centering
\includegraphics[width=0.72\linewidth]{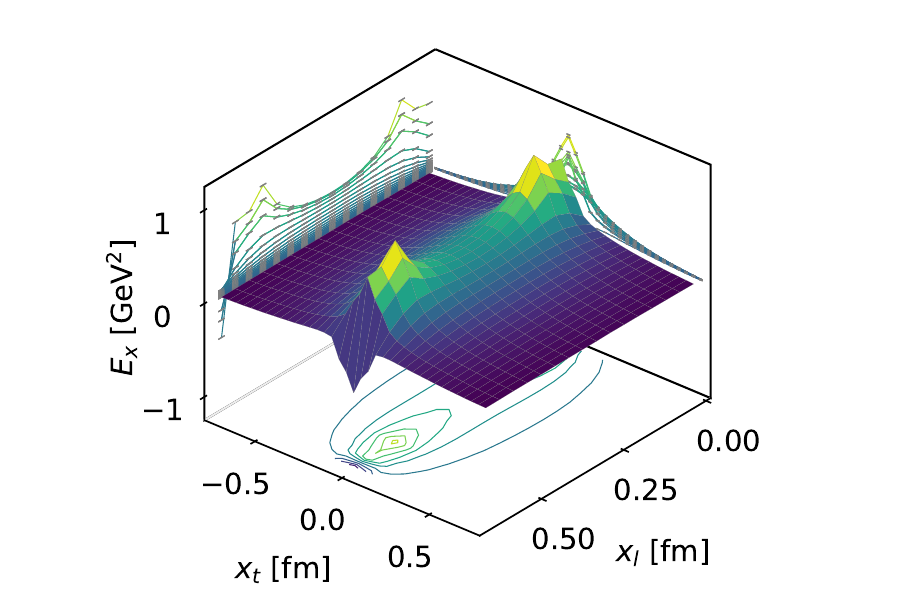} \\
\includegraphics[width=0.72\linewidth]{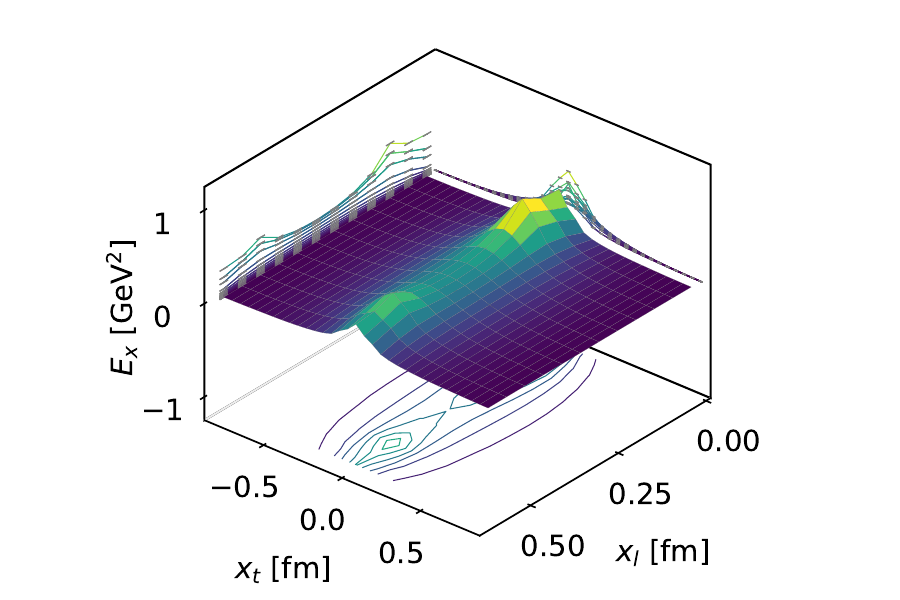}
\includegraphics[width=0.72\linewidth]{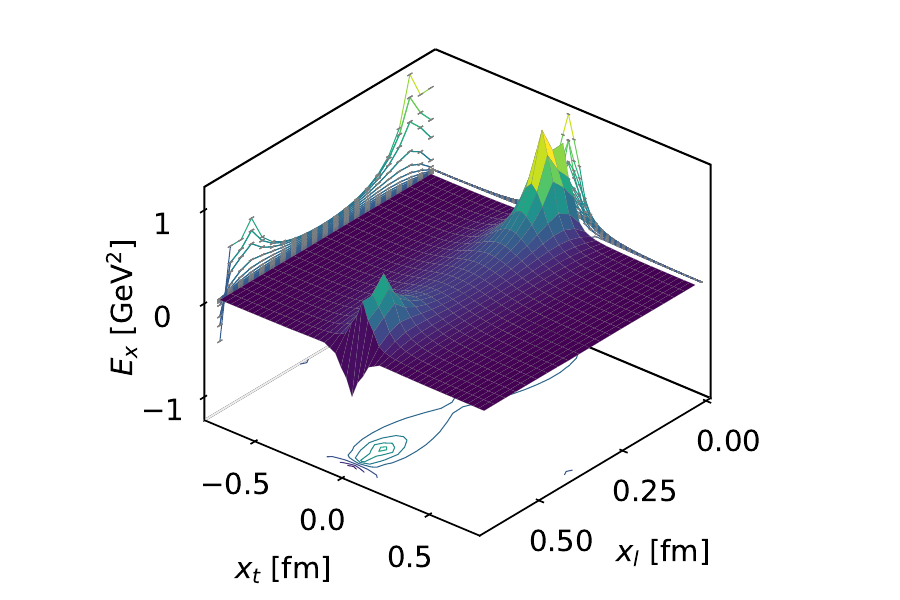}
\caption{3d plot of the full chromoelectric field for $T =0$, $1.2 T_\text{c}$, $2 T_\text{c}$ and $d=0.631$~fm.}
\label{fig:EX_FULL-3d}
\end{figure}

\begin {figure}[htb]
\centering
\includegraphics[width=0.72\linewidth]{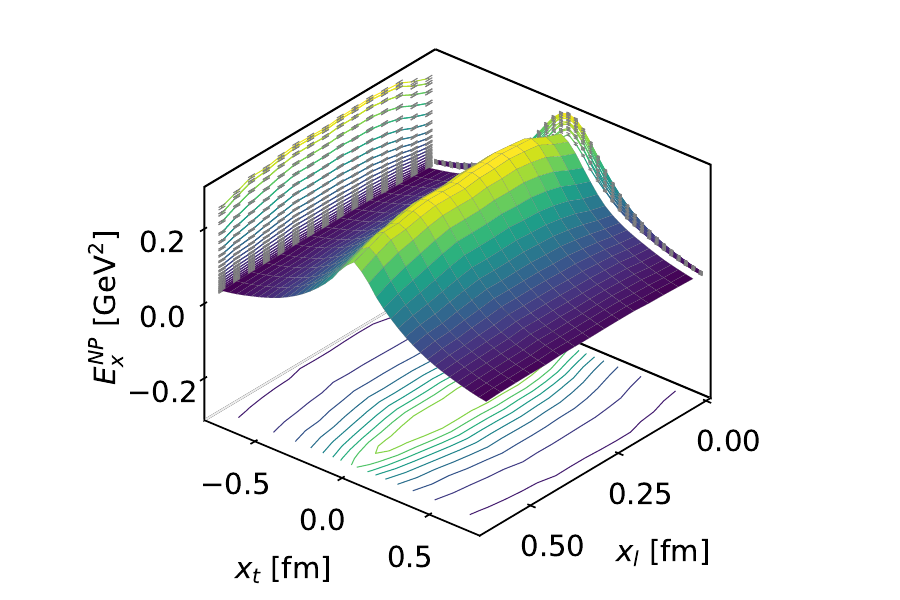} \\
\includegraphics[width=0.72\linewidth]{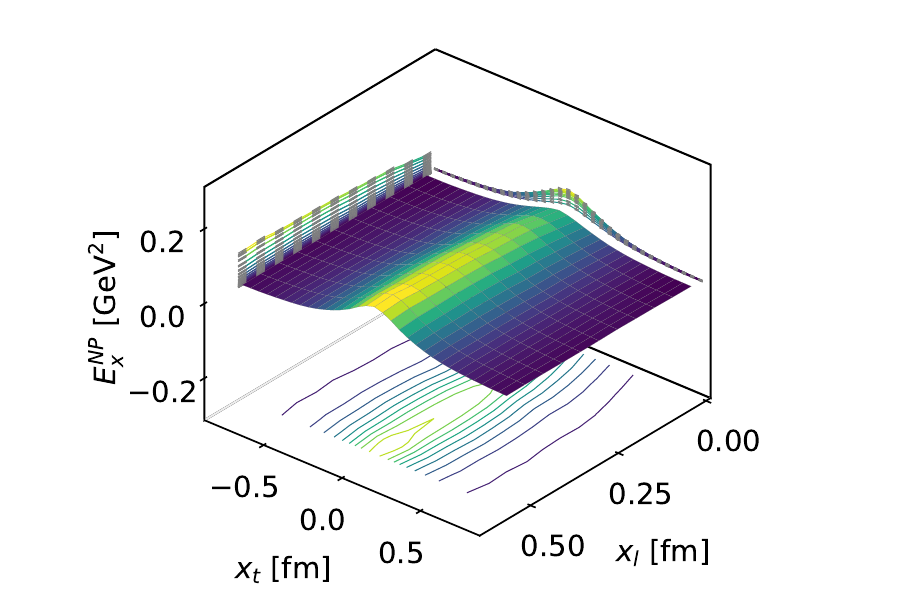}
\includegraphics[width=0.72\linewidth]{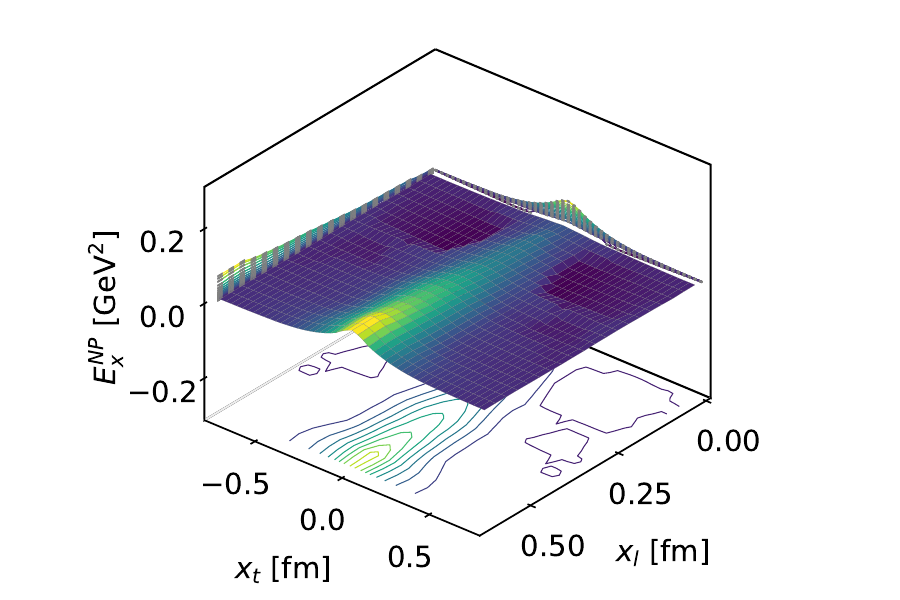}
\caption{3d plot of the nonperturbative chromoelectric field for $T =0$, $1.2 T_\text{c}$, $2 T_\text{c}$ and $d=0.631$~fm.}
\label{fig:EX_NP-3d}
\end{figure}

\begin {figure}[htb]
\centering
\includegraphics[width=0.72\linewidth]{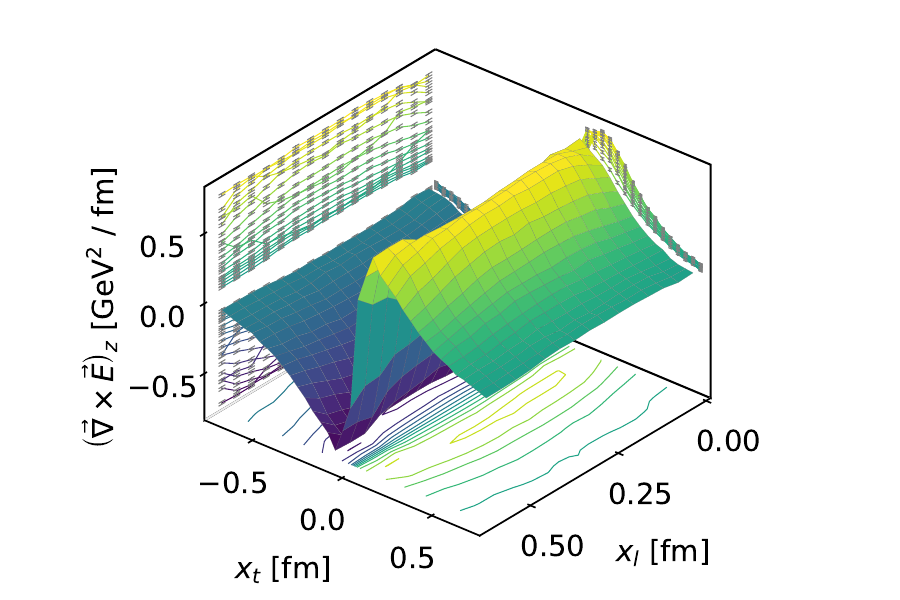} \\
\includegraphics[width=0.72\linewidth]{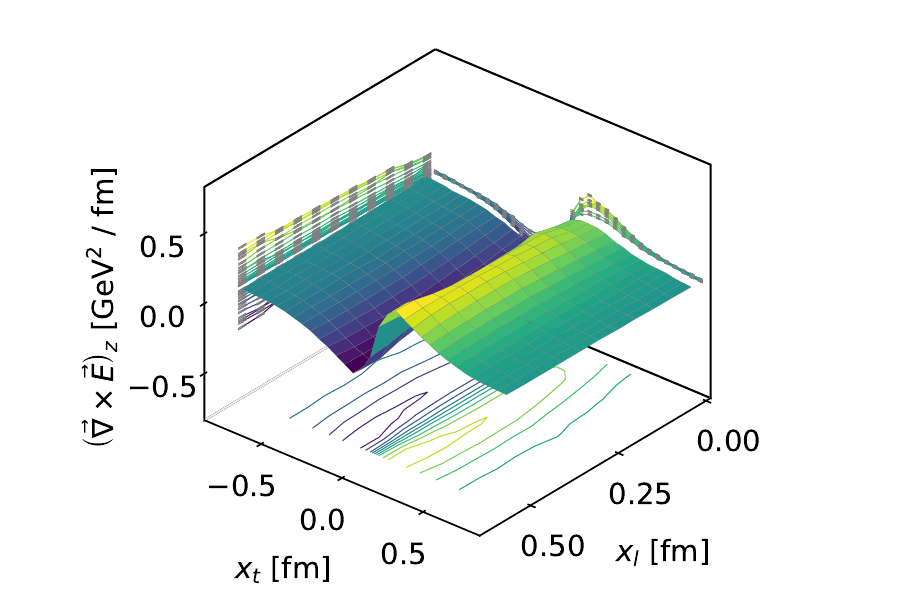}
\includegraphics[width=0.72\linewidth]{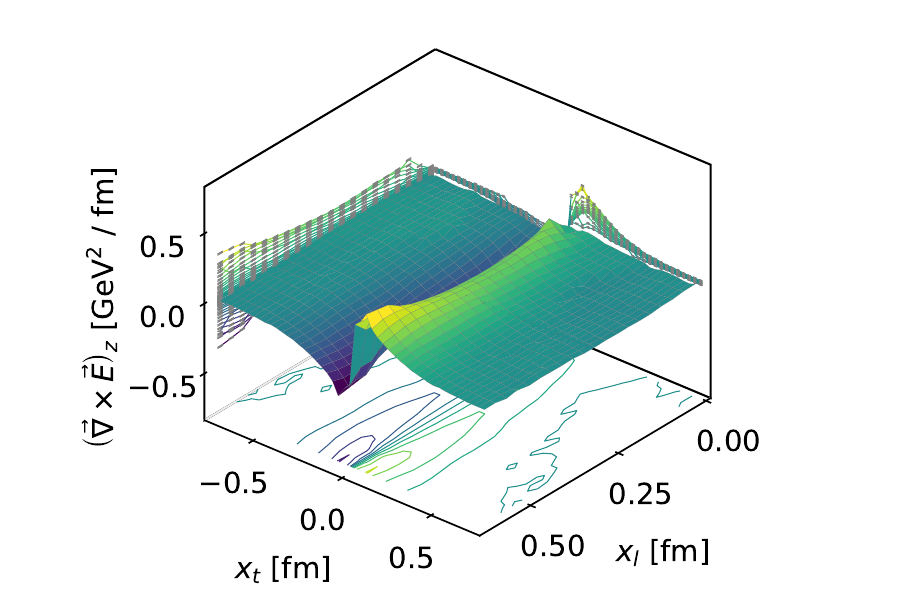}
\caption{3d plot of the magnetic current density for $T =0$, $1.2 T_\text{c}$, $2 T_\text{c}$ and $d=0.631$~fm.}
\label{fig:CURL_E-3d}
\end{figure}

\cref{fig:EX_FULL-3d,fig:EX_NP-3d,fig:CURL_E-3d} show the dependence of the full longitudinal chromoelectric field, the nonperturbative chromoelectric field, and the chromomagnetic current density on the position $(x_l, x_t)$ for three different values of temperature ($T=0$, $T=1.2 T_\text{c}$ and $T=2 T_\text{c}$) and for the same quark-antiquark distance $d \approx 0.63$~fm. 

One can see that the full field continues to form a tube-like structure well after reaching the deconfinement temperature. The remnants of the flux tube are visible also 
in the nonperturbative field and current density plots, despite the values becoming much smaller at higher temperatures. 

Another important observation is the lack of symmetry between the quark and antiquark on the
3d plots at high temperatures -- closer to the antiquark the full field values are much smaller, and the nonperturbative field and current density values are much larger than those close to the quark. The behavior of the full field suggests that the smearing required to perform the effective renormalization away from the quark at high 
temperatures is so large that the field is (partially) destroyed by smearing.

The growth of the nonperturbative field and current density suggests that our method of fixing the smearing amount (maximizing the signal value) might be inappropriate for very small signals -- at large distances we cannot distinguish the actual field value from the subtraction errors and end up overamplifying the latter.

These effects are much smaller near the midplane, so in what follows we will concentrate on the field at $x_l = d/2$. 

\subsection{Nonperturbative chromoelectric field}

\begin {figure}[htb]
\centering
\includegraphics[width=0.7\linewidth,clip]{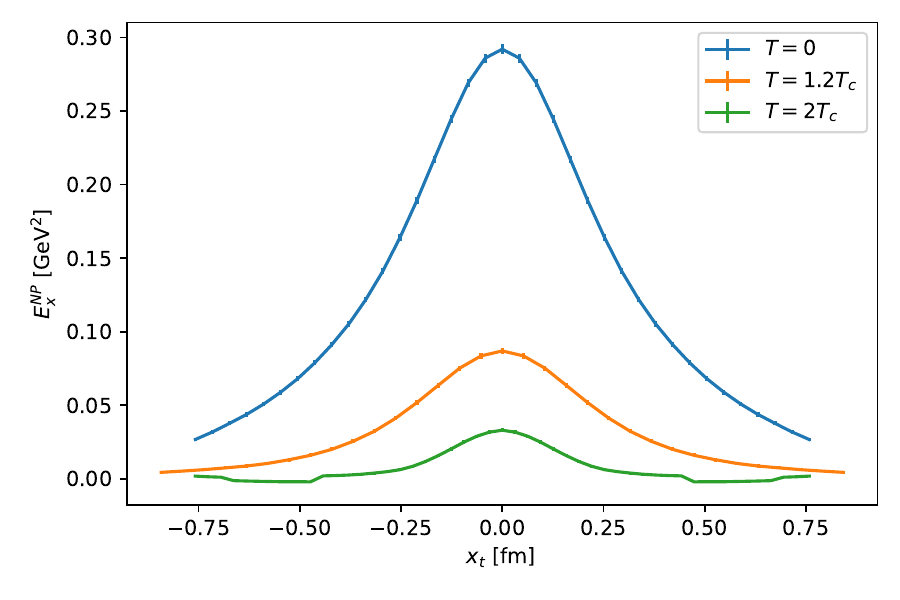}
\caption{The nonperturbative chromoelectric field at the midplane $x_l = d/2$, for $T =0$, $1.2 T_\text{c}$, $2 T_\text{c}$ and $d=0.632$~fm.}               \label{fig:nonperturbative_field}
\end{figure}

\cref{fig:nonperturbative_field} shows a midplane section of \cref{fig:EX_NP-3d}, providing a better view of the flux-tube remnant evaporation at $T > T_\text{c}$.

\begin {figure}[htb]
\centering
\includegraphics[width=0.7\linewidth,clip]{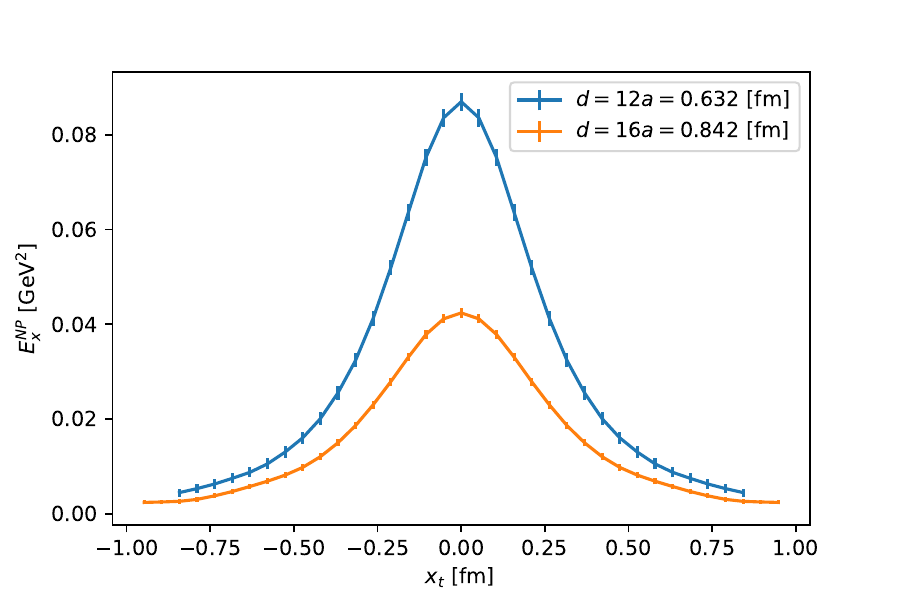}
\caption{The nonperturbative chromoelectric field at the midplane $x_l = d/2$, for $1.2 T_\text{c}$ and $d=0.632$, $0.842$~fm.}               \label{fig:nonperturbative_field-d_12_16}
\end{figure}

\cref{fig:nonperturbative_field-d_12_16} shows the values of the nonperturbative field at the same temperature $T = 1.2 T_\text{c}$, but for two distances $d = 0.632$~fm and $d = 0.842$~fm. One can see that when the quark-antiquark separation is increased by 1/3, the field values fall by more than 50 $\%$, and thus the flux-tube remnant does not create a linear potential at large distances. 

\subsection{Magnetic current density}

\begin {figure}[htb]
\centering
\includegraphics[width=0.7\linewidth,clip]{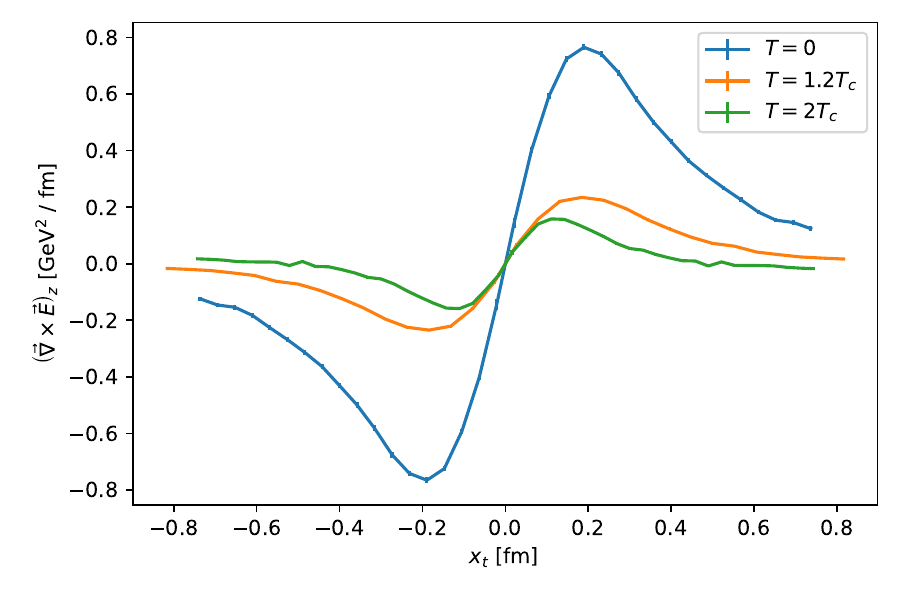}
\caption{Magnetic current density at the midplane $x_l = d/2$, for $T =0$, $1.2 T_\text{c}$, $2 T_\text{c}$ and $d=0.632$~fm.} \label{fig:current_density}
\end{figure}

\begin {figure}[htb]
\centering
\includegraphics[width=0.7\linewidth,clip]{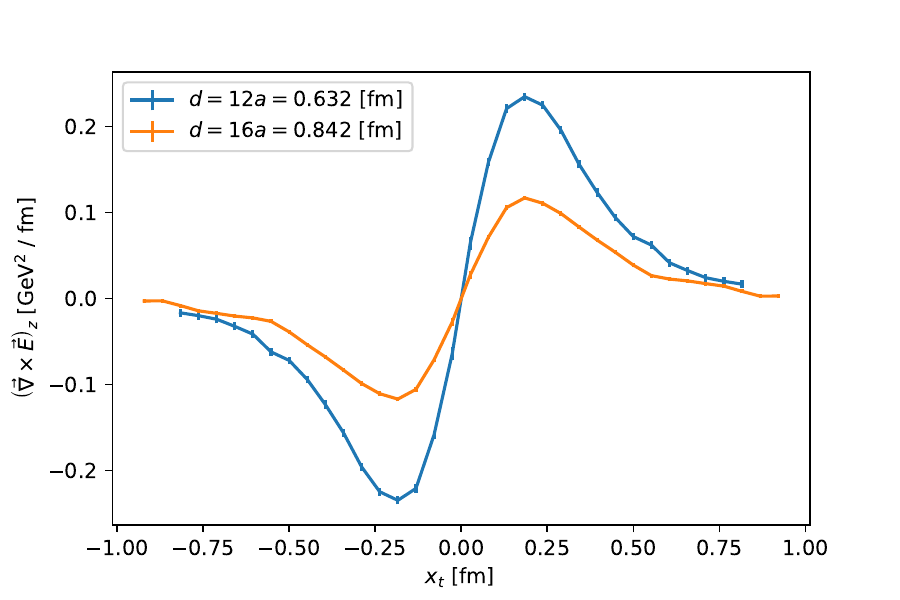}
\caption{Magnetic current density at the midplane $x_l = d/2$, for $T = 1.2 T_\text{c}$ and $d=0.632$, $0.842$~fm.} \label{fig:current_density-d_12_16}
\end{figure}

The same analysis can be done for the magnetic current density that should generate the flux tube. \cref{fig:current_density} shows that the current density drops significantly when temperature becomes larger than $T_\text{c}$, and \cref{fig:current_density-d_12_16} shows that the current density at the midplane drops when the distance between quark and antiquark increases. 

\subsection{Field integrals: string tension and confining force }

We also extracted the values of the integrals of the nonperturbative field, obtaining from them the string tension $\sigma$ and the confining force $F$, according to the following formulas: 
\begin{align}
\sigma &= \int d x_t^2 \ \frac{E_x^\mathrm{(NP)}(x_t)^2}{2} \ , \nonumber \\ 
F &= 2 \int_0^d d x_l \int_0^\infty d x_t \ E_x^\mathrm{(NP)}(x_l, x_t) J_z(x_l, x_t) \ .  
\label{field-integrals}
\end{align}

The evaluation of  Eqs.~\eqref{field-integrals} was done by doing a spline interpolation of the lattice data, and replacing the integration over the whole
 transverse plane by the integration over the circle $x_t < x_{t,\mathrm{max}}$. Note that Eqs.~\eqref{field-integrals} can also be used above the deconfinement transition as  evidence of the flux-tube dissolution. 

The integration results are collected in \cref{integrals}. The stochastic error estimates were obtained using the usual jackknife procedure. The systematic error estimates on $\sqrt{F}$ were obtained from comparing the integral in the range given in column $x_l$ (the region in which we have direct data), with the integral of the
extrapolated field values in the full range, and by considering the asymmetry of the obtained field, through the comparison of integrals over two halves of the region, (0 -- $d/2$) and ($d/2$ -- $d$).

One can see that below $T_c$ both $\sqrt{\sigma}$ and $\sqrt{F}$ values 
are stable under variation of $T$ and $d$ both with temperature and with $d$, and roughly compatible with each other 
(if we take into account systematic errors on $\sqrt{F}$). 
Once we enter  the deconfinement phase, both $\sqrt{\sigma}$ and $\sqrt{F}$
become drastically smaller, but do not go to zero.

Still, one can see that they are
also reduced when the temperature, and, more importantly, the distance $d$, grow. 
Obviously, in this case $\sigma$ cannot be treated 
as a string tension, since the assumption that the chromoelectric field profile 
does not depend on $x_l$ and $d$ is no longer valid. 
Thus, in the deconfined phase $\sigma$ and $F$ just serve as a measure of 
the residual field strength. 

\begin{table}[tb]
\setlength{\tabcolsep}{0.35pc}
\centering
\caption[]{
Summary of string tension and confining force results for the
lattice setups considered in this work. 
The first three lines (shaded gray) give, for the sake of comparison, the corresponding determinations at zero temperature, taken from Refs.~{\cite{Baker:2018mhw,Baker:2019gsi,Baker:2022cwb}}.}
\begin{tabular}{ccccccccc}
\toprule
$\beta$ & $d$ [fm] & $T/T_\text{c}$  & 
$x_{l}$ [fm] & $x_{t,\mathrm{max}}$ [fm] & 
$\sqrt{\sigma}$ [GeV] & $\sqrt{F}$ [GeV] \\
\midrule
\rowcolor{Gray}
\rule[-0.7em]{0pt}{1.9em}
6.240 & 0.511 & 0 &
0.032 -- 0.415 & 1.022 &
0.4742(15) & 0.4859(8)${}^{+645}$ \\
\rowcolor{Gray}
\rule[-0.7em]{0pt}{1.9em}
6.544 & 0.511 & 0 & 
0.021 -- 0.448 & 0.511 &
0.4692(23) & 0.5165(15)${}^{+611}_{-214}$ \\
\rowcolor{Gray}
\rule[-0.7em]{0pt}{1.9em}
6.769 & 0.511 & 0 &
0.016 -- 0.463 & 0.511 &
0.467(7) & 0.530(4)${}^{+547}_{-322}$ \\
\rule[-0.7em]{0pt}{1.9em}
6.554 & 0.631 & 0 & 
0.021 -- 0.610 & 0.736 & 
0.487(6) & 0.5635(23)${}^{+700}_{-830}$  \\
\rule[-0.7em]{0pt}{1.9em}
6.100 & 0.947 & 0.8 & 
0.039 -- 0.829 & 1.144 & 
0.535(20) & 0.664(9)${}^{+167}_{-227}$ \\
\hdashline
\rule[-0.7em]{0pt}{1.9em}
6.381 & 0.632 & 1.2 & 
0.026 -- 0.553 & 0.818 & 
0.129(4) & 0.1701(25)${}^{+125}_{-555}$ \\
\rule[-0.7em]{0pt}{1.9em}
6.381 & 0.842 & 1.2 & 
0.026 -- 0.816 & 0.921 & 
0.0733(25) & 0.1007(10)${}^{+284}_{-422}$ \\
\rule[-0.7em]{0pt}{1.9em}
6.544 & 0.631 & 1.5 & 
0.021 -- 0.610 & 0.736 & 
0.0625(19) & 0.0953(8)${}^{+290}_{-412}$ \\
\rule[-0.5em]{0pt}{1.5em}
6.248 & 0.632 & 1.5 &
0.032 -- 0.600 & 0.853 &
0.0556(8) & 0.1000(5)${}^{+302}_{-479}$ \\
\rule[-0.7em]{0pt}{1.9em}
6.778 & 0.632 & 2.0 & 
0.016 -- 0.616 & 0.742 & 
0.0305(14) & 0.0694(9)${}^{+248}_{-412}$ \\
\bottomrule
\end{tabular}
\label{integrals}
\end{table}

\section{Conclusions}

We have investigated, by Monte Carlo numerical simulations of  SU(3) pure gauge theory at
nonzero temperature, the behavior of the nonperturbative gauge-invariant longitudinal electric field, $\vec E^\text{NP}$, and of the magnetic current density,  $\vec J_{\rm mag}$, in the region between two static sources, a quark and an antiquark.

We have performed our numerical simulations for a range of values of the coupling where continuum scaling is satisfied and have considered four different temperatures, $T=0$,
$T=0.8 T_c$, $T=1.2 T_\text{c}$ and $T=2 T_\text{c}$. Most results are limited to the transverse plane midway between the sources, this choice being motivated by the observed sizeable asymmetry between the quark and the antiquark regions, probably due to the ineffectiveness of the smearing procedure in the lattice setup considered in this work, characterized by small values of the nonperturbative field.

Our findings can be summarized as follows. 

The full longitudinal chromoelectric field takes the shape of a flux tube even above the deconfinement transition; after subtraction of the perturbative component, the residual signal is still flux-tube like, but more and more suppressed as the temperature is increased above $T_\text{c}$. 

The flux tube remnant above $T_\text{c}$ does not generate a string tension.  The field drops considerably when the distance between the sources is increased, while keeping the temperature fixed, and the magnetic current circulating in the flux tube drops as well.

The breakdown of the Maxwell picture of the Yang-Mills flux tube is then a signal for the onset of deconfinement.

We plan to extend this analysis to  QCD (including quarks), and consider also the effect of nonzero baryon chemical potentials and other external sources.

\section*{Acknowledgements}
This investigation was in part based on the MILC collaboration's public lattice gauge theory code (\url{https://github.com/milc-qcd/}). Numerical calculations have been made possible through a CINECA-INFN agreement, providing access to HPC resources at CINECA. LC and AP acknowledge support from INFN/NPQCD project. FC and VC acknowledge support by the Deutsche Forschungsgemeinschaft (DFG, German Research Foundation) through the CRC-TR 211 ``Strong-interaction matter under extreme conditions'' -- project number 315477589 -- TRR 211. FC acknowledges the support by the State of Hesse within the Research Cluster ELEMENTS (Project ID 500/10.006).
This work is (partially) supported by ICSC – Centro Nazionale di Ricerca in High Performance Computing, Big Data and Quantum Computing, funded by European Union – NextGenerationEU.

\bibliography{qcd}

\end{document}